\documentclass[preprint,12pt]{elsarticle}

\usepackage{amssymb}
\usepackage{listings,xcolor}
\usepackage{float}
\usepackage{hyperref}
\usepackage{color,soul}
\usepackage{setspace}
\usepackage[utf8]{inputenc} 
\usepackage{soulutf8}
\usepackage{epigraph}
\usepackage{graphicx}
\usepackage{fancyhdr}

\journal{J Biomed Inform}

\begin{document}\sloppy

\spacing{1.5}

\begin{frontmatter}



\title{Semantic processing of EHR data for clinical research}
%
%



\author{Hong Sun\corref{cor1}}
\ead{hong.sun@agfa.com}
\cortext[cor1]{Corresponding author}

\author{Kristof Depraetere, Jos De Roo, Giovanni Mels, Boris De Vloed, Marc Twagirumukiza, Dirk Colaert\corref{cor2}}

\address{Advanced Clinical Applications Research Group, Agfa HealthCare, Moutstraat 100, 9000 Gent, Belgium }


\begin{abstract}
There is a growing need to semantically process and integrate clinical data from different sources for clinical research. This paper presents an approach to integrate EHRs from heterogeneous resources and generate integrated data in different data formats or semantics to support various clinical research applications. The proposed approach builds semantic data virtualization layers on top of data sources, which generate data in the requested semantics or formats on demand. This approach avoids upfront dumping to and synchronizing of the data with various representations. Data from different EHR systems are first mapped to RDF data with source semantics, and then converted to representations with harmonized domain semantics where domain ontologies and terminologies are used to improve reusability. It is also possible to further convert data to application semantics and store the converted results in clinical research databases, e.g. i2b2, OMOP, to support different clinical research settings. Semantic conversions between different representations are explicitly expressed using N3 rules and executed by an N3 Reasoner (EYE), which can also generate proofs of the conversion processes. The solution presented in this paper has been applied to real-world applications that process large scale EHR data.

\end{abstract}


\end{frontmatter}



\section{Introduction}
\label{intro}
After decades of development of electronic health records the integration of health records from different EHR systems has become a rising demand. The development of standard clinical information models is an attempt to tackle the storage and exchange of clinical data. Standards like HL7 \cite{hl7}, openEHR \cite{OpenEHR}, ISO 13606 \cite{iso13606}, etc. are developed to store or exchange patient records with structured formats. However, the semantic interoperability between different standards remains a challenge. 

In order to cope with semantic differences between EHRs we mapped data from different systems to semantic representations expressed with a core ontology \cite{ schober2010debugit} in the DebugIT project \cite{debugit}. Clinical data stored in different EHR systems are first mapped to semantic representations with local semantics of their respective EHR system and then mapped to expressions using the core ontology \cite{sun2012semantic}.

To further improve reusability, in the later SALUS project \cite{Salus}, we built semantic patterns for relevant clinical domains \cite{CREAM}, \cite{salus-d432} by reusing existing public ontologies and standard terminologies. We named such patterns Clinical Research Entity Advanced Model (CREAM) and aimed to achieve semantic interoperability between EHR systems and clinical applications by mapping their data to semantic expressions following CREAM.

The semantic interoperability achieved through a core ontology, or a harmonized domain information model such as CREAM, is nevertheless still fragile: interoperability is only achieved when the involved parties make their commitment to the common information model. Stakeholders of this common information model are not limited to the data providers (i.e. EHR systems) but also include data consumers (e.g. many clinical research applications). It is difficult to adapt research applications, which are built on top of a dedicated clinical data model, e.g. i2b2 \cite{ i2b2}, OMOP \cite{ OMOP}, to directly consume data expressed with domain semantics such as CREAM.

It is therefore important that data from a clinical data source can be mapped to a set of representations so as to achieve interoperability between the data source and multiple clinical research applications. This paper introduces a semantic data virtualization (SDV) scheme which is able to build multiple semantic data virtualization layers on top of a data source. Thus multiple clinical research applications can be supported. The SDV generates data in requested formats or semantics on demand. There is no need to dump the data to various representations upfront, thus avoiding the burden of synchronization with the source.

The SDV is constructed using RESTful services, which enables data transformation in a fully automated way. We use N3 rules to create the mappings between graph patterns expressed with different ontologies in different domains. As most of the existing N3 reasoners, e.g. EYE \cite{ de2013euler}, CWM \cite{ berners2000cwm}, etc. can generate a proof of a conversion process, we not only express the semantic conversions in an explicit way, but also the proof explains the executed conversion process. The SDV presented in this paper has been applied in the SALUS project to build the semantic interoperability layer. This paper generalizes the software components contained in the SDV from their specific implementation of the SALUS project, so that the SDV can be used in other projects.

The rest of this paper starts with discussions of related work. A summary of the different data layers in semantic processing of EHR for clinical research is presented, followed by an introduction of the architecture of the SDV and examples of mapping data between different layers by the SDV. Applications of the SDV in the SALUS and the AP-HP project are also given, together with a brief discussion regarding its performance.

\section{Related Work}
\label{sec:related-work}

To improve the interoperability of EHRs represented with different standards, mappings between different standards are developed and domain ontologies are created. Costa et al \cite{costa2011clinical} developed source ontologies for openEHR and ISO 13606, as well as a domain ontology which bridges the two standards. Data transformation is carried out through syntactic mapping between the archetypes of openEHR, ISO 13606 and the archetype model of the domain ontology. Gonçalves et al \cite{gonccalves2011using} build up a domain ontology of ECG and map schemas of three ECG standards directly to the domain ontology. They find it difficult to develop a domain ontology to which different standards can directly be mapped. 

Besides the report of directly mapping source data to a semantic representation with domain ontologies there are also proposals to achieve the mapping through multiple steps, which first map the EHR data with source semantics and later convert to representations with domain ontologies. Martínez-Costa et al \cite{ martinez2014improving} describe the data layers between data sources and end applications and state the necessary steps towards EHR semantic interoperability, as well as the challenges in implementing the steps. Berges et al \cite{ berges2012toward} first obtain the ontological representations of relational databases and later map the database ontologies to their canonical ontology. The canonical ontology presented in \cite{ berges2012toward} reuses existing medical terminologies such as LOINC and SNOMED. 

Different methods are proposed to represent the target clinical model. Early research relies on using a core ontology to represent the target clinical model \cite{debugit}, \cite{costa2011clinical}, \cite{ gonccalves2011using}. However, the ontology itself does not provide guidelines on how it can be used to represent target clinical models. Ontology content patterns, which guide and standardize the meaning of the content of clinical models, are proposed as a close-to-user representation to improve reusability \cite{CREAM}, \cite{ martinez2015semantic}, \cite{ presutti2008content}. The Clinical Information Modeling Initiative (CIMI) \cite{CIMI} proposes a set of modeling patterns, defined as clinical models, that can act as guidelines for the creation of ontology content patterns. The National Patient-Centered Clinical Research Network (PCORnet) also developed their Common Data Model (CDM) \cite{PCORnet-cdm} to map data from PCORnet partners to a common model.

Many of the above mentioned approaches have been applied in projects that target semantically processing of EHR data for clinical research. The DebugIT project \cite{debugit} maps EHR data from eight hospitals across Europe to representations with a core ontology for epidemiological research. Clinical questions are expressed with the DebugIT core ontology. A query generation service translates the clinical questions to corresponding SPARQL queries expressed with the source ontology, which are then executed on SPARQL endpoints at each site. A conversion service using an EYE reasoner maps the source data to expressions with the core ontology. The EHR data remain at the local hospitals and the outcomes of clinical questions are aggregated and displayed in the central dashboard. 

The Strategic Health IT Advanced Research Projects (SHARPn) \cite{ pathak2013normalization} develops their Clinical Entity Models (CEMs) as target model for EHR processing. Natural language processing (NLP) is used to process unstructured data. EHRs from two data sources, after NLP and normalization processing, are mapped to XML instances that conform to the CEM XSDs and stored in a central repository after anonymization. In their use case EHR data from 10,000 patients are used for diabetes mellitus research \cite{ rea2012building}.

The EHR4CR project \cite{ EHR4CR} aims at reusing EHR data for clinical research purposes. Two clinical data warehouses (CDW) are used as target models: an i2b2 data warehouse and a data warehouse with an EHR4CR specific schema. In their pilot application \cite{ doods2014piloting} each site established a CDW locally and used an ETL process to load the CDW with data from their respective EHR system. Seven sites use the EHR4CR database schema for their CDW, while the remaining four sites use i2b2. They are able to send one centrally created feasibility query and execute it at eleven sites to receive aggregated feasibility numbers with two different types of database schemas. Since both the SHARPn and the EHR4CR project store converted data in separate data warehouses, they have the burden to keep the data synchronized between the data source and their target data warehouse.

The SALUS project \cite{Salus} aims to create the necessary semantic interoperability infrastructure to enable secondary use of electronic health records by various clinical tools for proactive post market safety studies. The semantic interoperability layer of the SALUS project is constructed following the semantic processing scheme presented in this paper. It uses a set of semantic patterns, namely CREAM, to represent the target clinical model. The source data to CREAM conversion is carried out on demand at run-time, which avoids maintaining extra data stores for converted data. The pilot application is carried out on two EHR systems which contain 1 million and 10 million patients respectively. Six clinical applications developed by different partners are successfully executed on the converted data at both sites.

\section{Semantic Data Virtualization}
\label{sec:sdv}

The research presented in this paper aims to achieve semantic interoperability between data sources and clinical research applications through a data virtualization mechanism. This section first shows the data layers as well as the data flow in the proposed semantic framework. Then the architecture of the SDV is introduced. Examples of using the RESTful services of the SDV to implement the data flow are also demonstrated.

\subsection{Data Layers}
\label{sec:data-layer}

\begin{quotation} \emph{“The purpose of abstraction is not to be vague, but to create a new semantic level in which one can be absolutely precise.”}\\
\hfill ---Edsger Dijkstra
\end{quotation}

\begin{figure*}
\centering\includegraphics[width=0.55\linewidth]{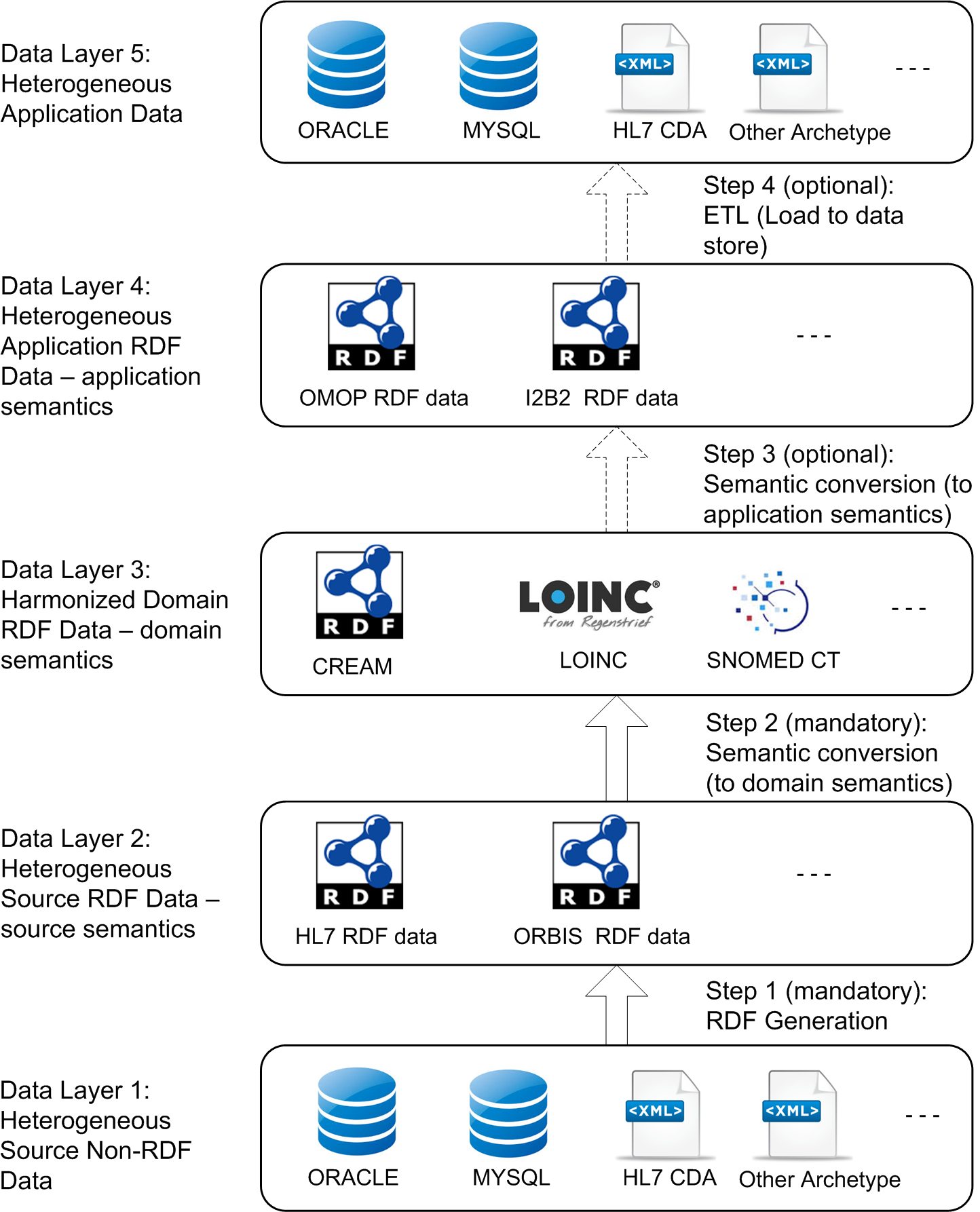}
\caption{Semantic data processing for clinical research}
\label{fig:data-flow}
\end{figure*}

In our previous work we used a two-step formalization approach, which first formalized operational data with its local semantics, and later converted the data with local semantics to data with domain semantics \cite{sun2012semantic}. Both local and domain semantics are precisely defined with ontologies. In this paper we extend this approach to further semantically process the data with domain semantics to application semantics, so as to achieve interoperability towards different clinical research applications. Figure \ref{fig:data-flow} shows the data layers in semantic data processing for clinical research as well as the needed actions to transfer data between the layers.

\paragraph{\textbf{Layer 1: Heterogeneous Source Non-RDF Data}}

On data layer 1 EHRs represented in different non-RDF formats are kept in their original formats in their respective data repositories. For example, a patient record may be stored as entries in tables of a relational database or as XML files in a data store. An excerpt of sample patient data in a relational database is shown in Table 1, Section \ref{sec:example-ddo}. Some EHR systems already established their own data warehouse where business intelligence is applied on their raw EHR data. Unless their data is stored in RDF format, those data warehouses are located together with their raw data counterparts on data layer 1. Compared with using a native EHR database, the advantage of using such a normalized data warehouse is that the stored data is normally pre-processed in the ETL (Extract-Transform-Load) process.  It is therefore normally more structured and easier to formalize. The disadvantage is that some context details stored in the daily practice database might be missing when the data is mapped to an abstracted data warehouse. One is free to choose the abstracted data or raw data, or even both, as their data source.

RDF generation (Step 1) is required to map non-RDF EHR data from their native representations (Data Layer 1) to their corresponding RDF representations (Data Layer 2). Although it is not common, an EHR data source can also be an RDF one, e.g. an RDF store. If a data source is already an RDF one, we consider it is already located in Data Layer 2. Then the RDF generation process (Step 1) can be skipped.

\paragraph{\textbf{Layer 2: Heterogeneous Source RDF Data}}

On data layer 2 heterogeneous source data are represented in RDF with source semantics (using source ontologies). An excerpt of sample patient data represented with source semantics is shown in Listing 2, Section \ref{sec:example-ddo}. The source ontologies are generated from source repositories respectively. For example, we generated a source ontology for a data source based on a one-to-one mapping from its database schema, following the policies stated in the W3C Recommendation of RDB2RDF direct mapping \cite{ world2012direct}:

\begin{itemize}
\item	a database table is mapped to an RDFS class (rdfs:Class); 
\item	a database table column is mapped to an RDF property (rdf:Property);
\item	the database data type of a field is mapped to the XSD data type range class of the property. One exception is that if a field is a foreign key, its range is the class that the foreign key points to.
\end{itemize}

The data on data layer 2, represented by the source ontology, still can not generally be understood by external systems. Semantic conversion (Step 2) is therefore required to convert such heterogeneous RDF data on data layer 2 into representations with harmonized domain semantics (Data Layer 3), so that it can be understood by external parties.

\paragraph{\textbf{Layer 3: Harmonized Domain RDF Data}}
On Data Layer 3 RDF data are harmonized and represented with domain semantics (using domain ontologies). An excerpt of sample patient data represented with domain semantics is shown in Listing 4, Section \ref{sec:example-do}. In order to maximize interoperability terminology mappings should be carried out in this step as well. For example, lab test results expressed with local lab codes are mapped to expressions with LOINC codes. The code mapping process can be carried out in an automated and semantic way if mappings between the local coding system and the standard coding system exist \mbox{\cite{hussain2014justification}}. Since high quality mappings are scarce labor intensive work is often required to create such mappings. Nevertheless, the created mappings can be reused in other projects. Once the local codes are mapped to the standard ones, it is possible to further map them to other codes with existing mappings \mbox{\cite{hussain2014justification}}.

It is advised by \cite{ heath2011linked} to reuse existing vocabularies wherever possible, rather than reinvent, so as to maximize data interoperability. We share the same view and we deem that the task of building a data layer with harmonized domain semantics is not to create a new ontology to serve as an interoperability hub inside one single project. Rather it is about reusing existing open standard ontologies in the relevant domain, so as to generate domain data which are interoperable in the related community and even across different projects \cite{martinez2015semantic}, \cite{ presutti2008content}. We therefore build a set of semantic patterns (CREAM \cite{CREAM}) to cover relevant parts of the clinical domain.

Semantic interoperability originating from different data sources is achieved on data layer 3. Data represented with harmonized domain semantics can be aggregated together and can be consumed by different clients for clinical research. If the semantics that a clinical research application relies on is already defined in the domain, such an application can directly consume data from layer 3. Otherwise a semantic conversion (Step 3) is required to convert the RDF data with harmonized domain semantics to representations with application semantics.

\paragraph{\textbf{Layer 4: Heterogeneous Application RDF Data}}

On data layer 4 RDF data are represented with heterogeneous application semantics (using application ontologies). An excerpt of sample patient data represented with application semantics is shown in Listing 6, Section \ref{sec:example-ao}. Application ontologies are generated from the data model or database structure of target applications. The generation of the application ontology is recommended to follow the same guidelines that we used to create the source ontology on layer 2.

Applications that are capable of processing RDF data can consume RDF data from layer 3 or layer 4. For applications that are built on particular archetypes or database structures (e.g. i2b2, OMOP, etc.) a lightweight ETL job (Step 4) is required to load data into the target data repositories. The ETL job is considered lightweight because most of the transformation process is already performed.

\paragraph{\textbf{Layer 5: Heterogeneous Application Data}}

On data layer 5 processed data are finally stored in different clinical research data repositories to support their corresponding clinical research applications. An excerpt of sample patient data in the OMOP database is shown in Table 2, Section \ref{sec:example-ao}.

\bigskip

In practice it is also possible to merge step 1 and step 2 to directly map data on Layer 1 to a semantic representation on layer 3. However, we would strongly recommend not doing so for the following reasons. 
\begin{itemize}
\item	Explicitness and provability. The two step conversion allows to express the semantic mapping logic explicitly in N3 rules, which enable to generate a proof when it is executed by an N3 reasoner (see Section \ref{sec:proof}). While the one step approach probably hard codes the mapping logic in an RDF generator, e.g. embedding SQL query templates in R2RML \cite{world2012r2rml}.
\item	Expressiveness. The two step approach can use complex conversion rules and advanced functions. While the expressions of the one step approach are limited. E.g. although R2RML can process complex expressions for retrieving source data via a SQL template, it is not able to generate complex expressions as its target.
\item	Performance. Mapping data to a semantic representation is not an easy task. The existing RDB-to-RDF mapping tools (e.g. a SPARQL endpoint) already exhibit performance issues (e.g. performance degradation with the use of optional statements)\cite{holford2012semantic}. Adding complex semantic mapping logic to the semantic mapping tool would further jeopardize its performance \cite{sun2012semantic}. The RDB to RDF implementation report \cite{r2r-implementation} also shows that there are less tools supporting a full fledged R2RML \cite{world2012r2rml} compared to the simplified direct mapping \cite{ world2012direct}. For example, the popular RDB to RDF tool D2RQ \cite{ bizer2004d2rq} only supports the direct mapping.
\end{itemize}

The process of loading harmonized RDF data (on layer 3) to heterogeneous data repositories (on layer 5) can be considered as reverse to the action of generating harmonized RDF data from heterogeneous data sources. It is also possible to merge step 3 and step 4 to load harmonized data on layer 3 to data repositories directly. We separate it into two steps for a similar reason when we normalize the data: it makes the conversion process explicit and also simplifies the ETL job. 

For some clinical research applications a set of ETL implementations has been developed to map data from a number of source databases to a dedicated clinical research database as e.g. OMOP CDM \cite{omop-cdm}. We consider these implementations as customized end-to-end mappings from data layer 1 to data layer 5 lacking reusability. Our implementation intends to build a bridge to map data with common domain semantics on data layer 3 to representations with application semantics on layer 5, which can be reused by different data sources. We do not prohibit the use of existing ETL implementations. But we consider applying such ETL implementations to be out of the scope of the proposed semantic data virtualization scheme.

It is also important to point out that the definitions of the layers are depending on the application setting. For example, in one project a clinical application is built on the OMOP database. The OMOP database is then considered as containing application data, which resides on data layer 5. Source data from other databases need to be transferred to the OMOP database through the SDV. In another project it is possible that the source data is stored in an OMOP database and it needs to be mapped to an i2b2 database for i2b2 based clinical applications. In the latter case the OMOP database is then considered as containing source data, which resides on data layer 1.

\subsection{Architecture of the Semantic Data Virtualization Solution}

\begin{figure*}
\centering\includegraphics[width=0.9\linewidth]{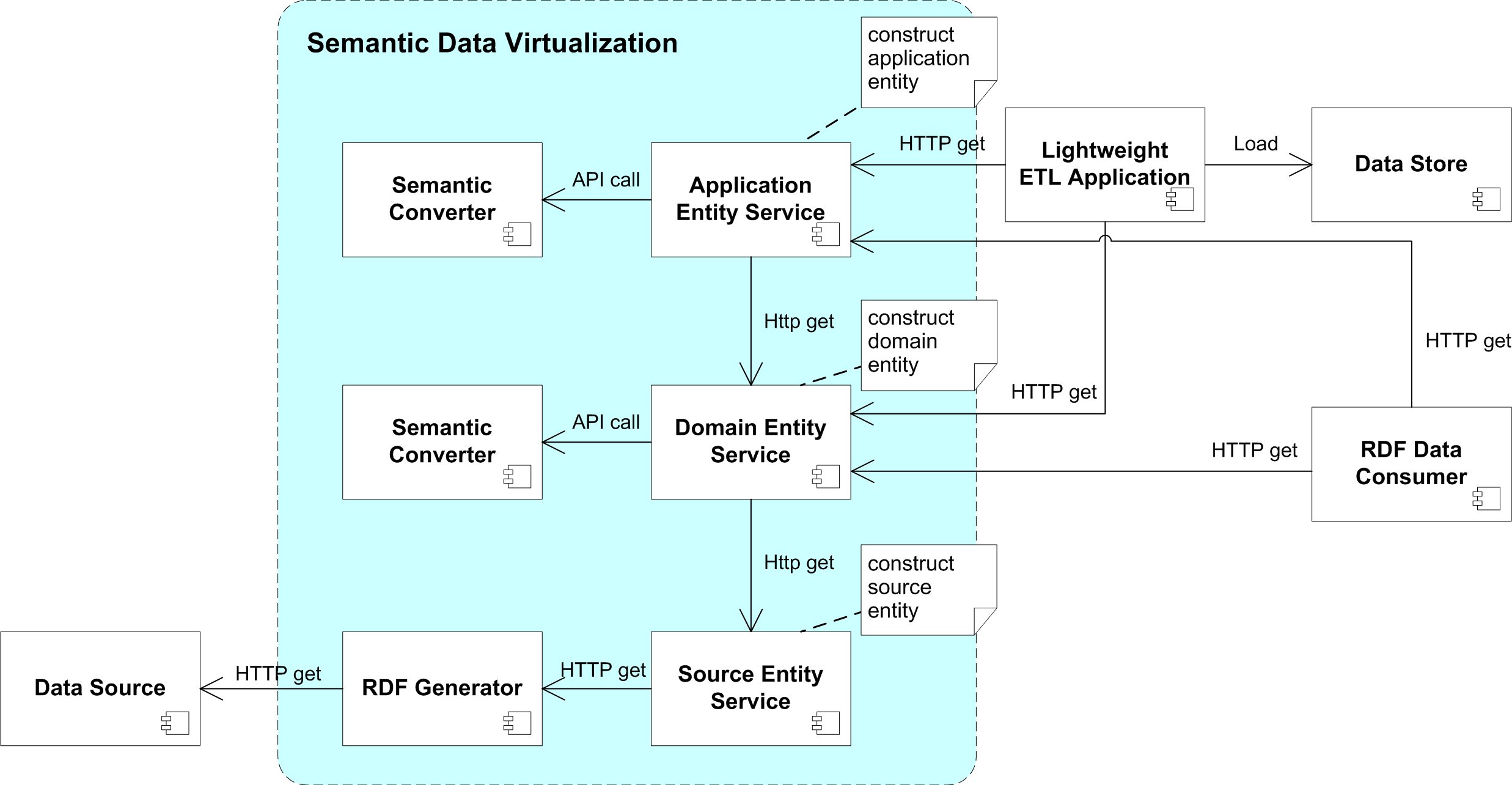}
\caption{Architecture of the Semantic Data Virtualization Solution}
\label{fig:architecture}
\end{figure*}

Figure \ref{fig:architecture} shows the architecture of the SDV solution, which implements the data flow shown in Figure \ref{fig:data-flow}. The RDF generator generates RDF data represented with source semantics (Data Layer 2 in Figure \ref{fig:data-flow}). If a data source stores non-RDF data, the RDF generator needs to map the content to an RDF representation (Step 1). The semantic converter reads semantic data in and converts its semantics. For example, it converts semantic data with source semantics to domain semantics (Step 2) or from domain semantics to application semantics (Step 3). The lightweight ETL task loads data expressed with application semantics to application repositories.

In our previous projects we found that, although we are able to convert data across different domains, it is extremely difficult, if not impossible, to translate queries among different domains. For example, a user could write a SPARQL query (using the domain ontology) to retrieve relevant data from the domain layer (providing the data is already expressed with domain semantics). However, it is difficult to translate this domain query to a set of queries expressed with the relevant data source ontology, extract data from the targeted data source, convert the data to expressions with domain semantics, and finally provide an answer to the domain query. In the DebugIT project we used a query generation tool to generate SPARQL queries expressed in source ontology from a SPARQL query expressed in domain ontology. The generated ones are then sent to local SPARQL endpoints and mapped to SQL queries by D2RQ. We found it difficult to develop such a query translation tool which is fully independent of the data sources and conversion rules. Additionally, some interpreted SPARQL queries also experienced performance issues. Lastly, it is also difficult to construct relevant conversion rules to translate the extracted data to the domain semantics for an open domain query. In the end we created a query generation tool that only supports a limited set of query templates.

It is however possible to upfront convert the entire data source to the domain semantics and store this converted data in a dedicated RDF store. The drawback of this approach is that it adds an extra burden on maintaining the additional data store. In particular, it is difficult to manage (timely) updates in case of changes in the data source, unless the data source supports some kind of data change notification mechanism. Moreover, if there are multiple applications (using different application ontologies) built on top of one data source, this approach requires building and maintaining dedicated RDF data stores for each of these applications.

We thus decide to keep the source data in their original data store (Layer 1 in Figure \ref{fig:data-flow}) and build virtual layers on top of the source data. Different from our previous implementation in the DebugIT project, we now group relevant information resources together as an RDF graph pattern (example see Listing 1). We call the representations of these resources "Entities". An entity can be located on Layer 2, 3 and 4. (See Figure \ref{fig:data-flow}). Our CREAM specification currently defines 19 entities in the relevant clinical domain, ranging from general ones, such as patient demographics, to specific clinical objects like immunotherapy. We now choose to use the REST architectural style for defining the entities as well as the architecture of the SDV. REST has the following favorable constraints: client-server, stateless, cacheable, layered system, uniform interface and (optionally) code on demand. 

We do not support a data consumer to send its own custom query to the SDV. A user is mandated to communicate with the SDV with HTTP requests through the HTTP API interface in the REST style. Entities are produced by RESTful services. Actions such as retrieving or converting data are handled by these RESTful services, following the configurations deployed at deployment time. A user may retrieve entities from the SDV with HTTP requests. 

The key components of the SDV are as follows:

\paragraph{\textbf{RDF Generator}}

An RDF generator acts as an interface to provide RDF data that is represented with source semantics (Data Layer 2). For those non-RDF data sources, their RDF generator should implement the RDF generation process (Step 1 in Figure \ref{fig:data-flow}) so as to transfer the non-RDF data to its corresponding RDF representation. In our implementation we build a SPARQL endpoint as the RDF generator for our Agfa HealthCare ORBIS EHR system. The ORBIS EHR system stores its data in an Oracle database. We generate an ORBIS ontology from the ORBIS database schema following a 1-1 mapping policy (see Section \ref{sec:data-layer}, Layer 2). The endpoint acts similar to D2RQ \cite{ bizer2004d2rq}: it translates a SPARQL query expressed with the ORBIS ontology into a SQL query and converts the returned SQL result set into RDF. There is no obligation to use a SPARQL endpoint as the RDF generator. Any tool that is able to generate the requested RDF data from the data source could be used as an RDF generator.

\paragraph{\textbf{Source Entity Service}}

A source entity service provides an interface to call the RDF generator to construct source entities. A source entity is a set of relevant RDF data, which resides on data layer 2. A sample source entity is shown in Listing 2. Each source entity is assigned a URL which is distinct from other entities. The URL is mapped to a corresponding configuration folder containing a query template (i.e. Listing 1) to instruct the RDF generator how to retrieve data.

The source entity service allows to extract RDF data from a data source in REST style. Once an HTTP GET request is sent to the source entity service (see Section \mbox{\ref{sec:example-ddo}}), a query will be generated using the query template together with query parameters specified in the request. The RDF generator will execute the query and return the resulting source entity via the HTTP API.

\paragraph{\textbf{Semantic Converter}}

A semantic converter takes one or more entities as input data and converts them to a new entity with different semantics. The semantic converter can generate domain entities on data layer 3. It is also capable of generating application entities as described on data layer 4. The input data (entities) of a semantic converter can be either source entities from RDF generators or domain entities from other semantic converters. There are a few approaches to fulfill the request. By using N3 rule engines as CWM, EYE, one can express the requested conversion with N3 rules following the pattern \{input-graph\} =$>$ \{output-graph\}. It is also possible to realize the conversion with a SPARQL query following the pattern CONSTRUCT \{output-graph\} WHERE \{input-graph\}. TopBraid Composer provides a visual mapping tool SPINmap \cite{SPINMap} to create SPARQL mapping rules.

In our implementation we use EYE as the semantic converter to execute conversion rules expressed in N3. EYE is an open source N3 reasoning engine which translates Notation 3 into Prolog. It is well maintained, fast in reasoning speed and can also generate proofs of the reasoning process to build additional trust. EYE supports a list of builtins \mbox{\cite{eye-builtins}}, which includes all Prolog builtins. It therefore has the expressive power of Prolog, which is Turing complete. One can write N3 programs having a NEXPTIME complexity but for all our applications, our validation checks that the worst case time and space complexity is n*log(n).

\paragraph{\textbf{Domain Entity Service}}

A domain entity service provides an interface to call the semantic converter to generate domain entities. A domain entity is a set of relevant RDF data, which resides on data layer 3. A sample domain entity is shown in Listing 4. Similar to the source entity, each domain entity is identified with a URL which is distinct from the others. Each domain entity has a corresponding configuration folder which contains a set of conversion rules (i.e. Listing 3) and optionally a query rule to filter the output result. The needed input data can either be contained in the configuration folder or specified as a URL query parameter and be retrieved at run-time. Both source entities and domain entities can be used as input data. 

Once an HTTP GET request is sent to the domain entity service (see Section \mbox{\ref{sec:example-do}}) the service will parse the request and send requests to retrieve either source entities or domain entities as input data. The semantic converter could request multiple input data from diverse data sources. Once the requested data are received, the EYE reasoning engine (semantic converter) will convert the input data with the stated conversion rules.

\paragraph{\textbf{Application Entity Service}}
An application entity service provides an interface to call the semantic converter to generate application entities. An application entity is a set of relevant RDF data, which resides on data layer 4. A sample application entity is shown in Listing 6, and a sample HTTP GET request for an application entity is shown in \mbox{\ref{sec:example-do}}. The application entity service implementation is the same as the implementation of the domain entity service. The only difference between a domain entity service and an application entity service is their configuration. The former one maps data to representations with domain semantics (Data Layer 3) while the latter one maps data to representations with application semantics (Data Layer 4).

\paragraph{\textbf{Lightweight ETL Application}}
A lightweight ETL application retrieves data from either application entity service or domain entity service and loads them to target data repositories (e.g. i2b2, OMOP, etc.). Since most of the data extraction and transformation effort is already done by the domain or application entity services, the ETL job does transformations on a minimal level and mainly focuses on data loading.

\bigskip

\begin{figure}
\centering\includegraphics[width=0.6\linewidth]{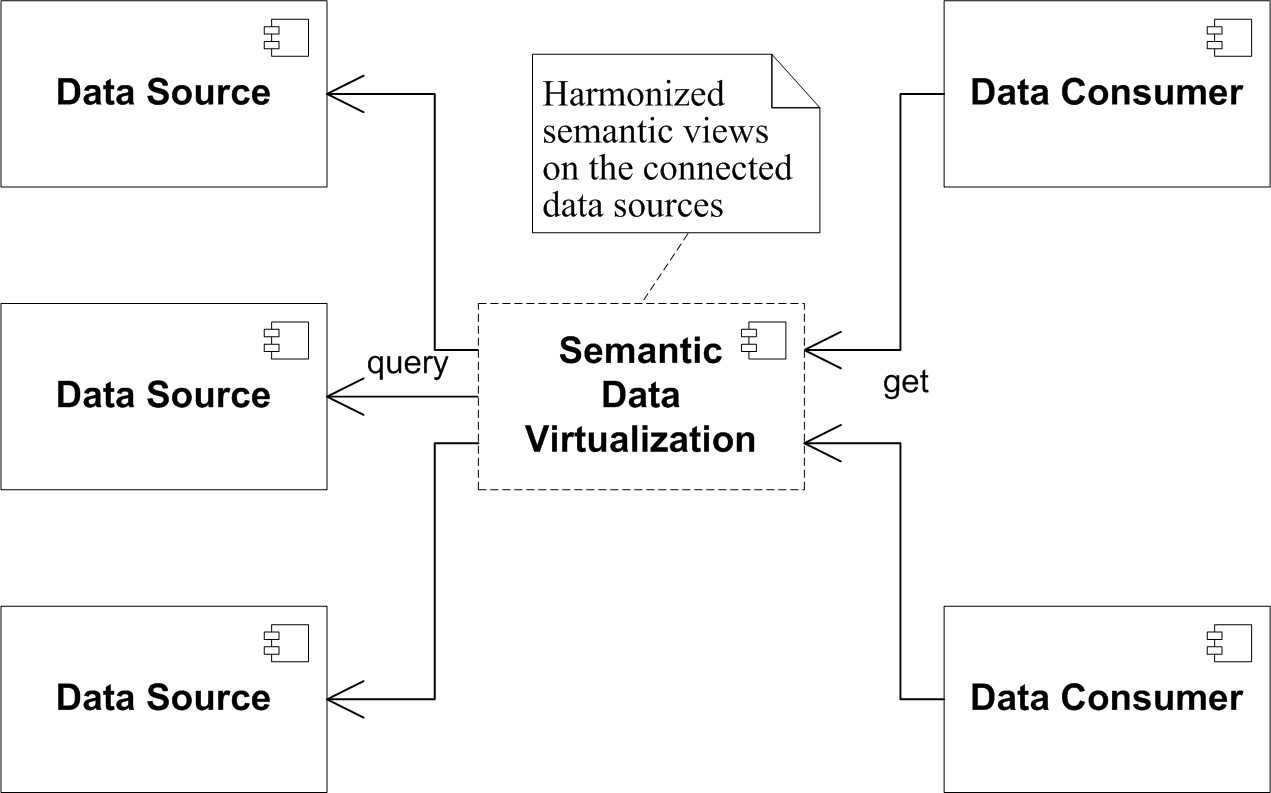}
\caption{Semantic interoperability by SDV}
\label{fig:interoperability}
\end{figure}

Semantic interoperability between data sources and clinical research applications can be achieved through semantic data virtualization: data from independent data sources are converted into a harmonized view (Data Layer 3), so that data from different data sources are integrated. To fulfill the request of a data consumer, the SDV can also translate data represented with domain semantics to data represented with the semantics of target application, and thus provide semantic interoperability towards these applications. Data sources such as EHR data, healthcare and life science data can for example be integrated by this mechanism to expand the knowledge domain. The hub and spoke architecture presented in Figure \ref{fig:interoperability} shows that a data consumer can consume data from multiple data sources and a data source can provide data to multiple data consumers.

\subsection{Example}

This section uses an excerpt of the patient demographics entity to demonstrate how the SDV generates different representations of data through its RESTful entity services. Proofs of reasoning processes, which can be used to build trust of the conversion, are also introduced at the end of this section. The sample entities and conversion rules used in this section, as well as the corresponding proofs can be found online \cite{github-sample}.

\subsubsection{Retrieve Source Entity through Source Entity Service}
\label{sec:example-ddo}

The sample URL specified in (1) shows an example of how to retrieve a source entity. In the specified source entity, \emph{orbis/demographics/demographics}, \emph{orbis} denotes the data source (ORBIS EHR system), the first \emph{demographics} denotes the domain and the latter one denotes the requested source data. In our practice, the demographics domain entity requests both demographics (\emph{demographics/demographics}) and address (\emph{demographics/address}) data from our ORBIS data source as inputs. The sample URL shown in (1) calls the source entity service to retrieve demographics data of a specified patient. A patient URI (e.g. \emph{http://example.org/resource/Patient/1001}) can be used in the place holder as a filtering condition.

\begin{lstlisting}
(1): http://example.org/rdf_generator/entities/orbis/demographics/demographics?patient_uri={patient_uri_place_holder}
\end{lstlisting}

\begin{lstlisting}[float,frame=single,caption=Sample SPARQL query template for demographics source entity]
PREFIX patient: <http://www.agfa.com/orbis-schema/Patient#>
PREFIX natperson: <http://www.agfa.com/orbis-schema/Natperson#>
1 CONSTRUCT {
2	?patient patient:persnr ?person.
3	?person natperson:vorname ?vorname.
4	?person natperson:name ?name.
5	?person natperson:gebdat ?birthDateTime.
6 } WHERE {
7	?patient patient:persnr ?person.
8	OPTIONAL {?person natperson:vorname ?vorname.}
9	?person natperson:name ?name. 
10	?person natperson:gebdat ?birthDateTime. 
11	$if(patient_uri)$
12		FILTER (?patient = <$patient_uri$>)
13	$endif$ 
14 }
\end{lstlisting}

Listing 1 shows an excerpt of the SPARQL query template for the demographics source entity. The ontology used in the sample SPARQL query is generated from the database schema of the ORBIS EHR system. Once a request as (1) is received, the patient URI will be applied as the filter condition of the template in Listing 1 to construct a SPARQL query. The constructed SPARQL query is then executed by our ORBIS SPARQL endpoint. The SPARQL endpoint translates the SPARQL query into SQL statements, executes it on the database, and converts the returned SQL result set to RDF using the ORBIS ontology. Table 1 shows an entry of a sample record of the NATPERSON table in the ORBIS database. Listing 2 shows part of the result by calling the URL in (1) through the source entity service. It is obvious that the data represented with the source ontology (ORBIS ontology) is difficult to be understood by external bodies. Its interoperability potential is therefore very low.

\begin{table}[float, ht]
\caption{Sample data of NATPERSON table in ORBIS database (Data Layer 1)}
\centering
\begin{tabular}{l l l l}
\hline
\textbf{id} & \textbf{vorname} & \textbf{name} & \textbf{gebdat}\\
\hline
1001 & Agfa & Healthcare & 1990-02-08 00:00:00 \\
\hline
\end{tabular}
\label{tab:orbis-natperson}
\end{table}

\begin{lstlisting}[frame=single,caption=Sample source entity (Data Layer 2)]
<http://example.org/resource/Patient/1001> 	patient:persnr <http://example.org/resource/Natperson/1001>.
<http://example.org/resource/Natperson/1001> natperson:vorname "Agfa".
<http://example.org/resource/Natperson/1001> natperson:name "Healthcare".
<http://example.org/resource/Natperson/1001> natperson:gebdat "1990-02-08T00:00:00+01:00"^^xsd:dateTime.
\end{lstlisting}

\subsubsection{Retrieve Domain Entity through Domain Entity Service}
\label{sec:example-do}

The sample URL specified in (2) calls the domain entity service to retrieve a domain entity (demographics). It also uses a patient URI as the filter to restrict the result. Once the URL in (2) is called, it would first call the source entity service to retrieve the requested input data. This is done by calling the URL as described in (1) to retrieve the \emph{orbis/demographics/demographics} source entity, together with a separate call to retrieve the \emph{orbis/demographics/address} source entity. Once the data is retrieved, the source to domain conversion rules, as specified in Listing 3, are applied to generate the demographics domain entity for the specified patient. An excerpt of the generated domain entity is shown in Listing 4.

\begin{lstlisting}
(2): http://example.org/semantic_converter/entities/demographics?patient_uri={patient_uri_place_holder}
\end{lstlisting}

\begin{lstlisting}[float,frame=single,caption=Sample conversion rules from source to domain]
PREFIX schema: <http://schema.org/>
PREFIX patient: <http://www.agfa.com/orbis-schema/Patient#>
PREFIX natperson: <http://www.agfa.com/orbis-schema/Natperson#>
1    {	?patient patient:persnr ?person.
2     } => {
3	?patient a <http://snomed.info/id/116154003>. #Patient
4	?patient a schema:Person. }.
5
6    {	?patient patient:persnr ?person.
7	?person natperson:gebdat ?birthDate.
8    } => {
9	?patient schema:birthDate ?birthDate. }.
10
11  {	?patient patient:persnr ?person.
12	?person natperson:name ?familyName.
13   } => {
14	?patient schema:familyName ?familyName. }.
\end{lstlisting}

Listing 3 shows a set of sample conversion rules, which convert the ORBIS demographics source entity (Listing 2) to a domain entity (Listing 4). '=$>$' stands for log:implies \cite{log}, its subject (the left side graph of '=$>$') is the antecedent graph, and the object (the right side graph) is the consequent graph. The first rule (Line 1-4) states a patient instance (in the ORBIS data source) is an instance of the SNOMED CT concept 116154003 (Patient), and an instance of the class schema:Person in the domain, as defined by CREAM. The other two rules translate birth date and family name from expressions with the source ontology to expressions with the domain ontology. It is expected that the domain entity presented in Listing 4 is easier to be interpreted by external bodies compared to its source entity counterpart in Listing 2.

\begin{lstlisting}[float,frame=single,caption=Sample domain entity (Data Layer 3)]
<http://example.org/resource/Patient/1001> a schema:Person,
	<http://snomed.info/id/116154003>. 
<http://example.org/resource/Patient/1001> schema:familyName "Healthcare".
<http://example.org/resource/Patient/1001> 
	schema:birthDate "1990-02-08T00:00:00+01:00"^^xsd:dateTime.
\end{lstlisting}

\subsubsection{Retrieve Application Entity through Application Entity Service}
\label{sec:example-ao}

As shown in Figure \ref{fig:architecture}, domain entities can be further converted to application entities, so that a lightweight ETL process can be performed to load the resulting RDF data to a target clinical database (e.g. OMOP, i2b2, etc.) for dedicated clinical research. The sample URL specified in (3) calls the OMOP application entity service to generate an OMOP demographics entity for a patient. It first retrieves the domain demographics entity as input data. This is done by calling the URL as specified in (2). Then it applies the domain to application conversion rules (Listing 5) to generate the application entity (Listing 6).

\begin{lstlisting}
(3): http://example.org/semantic_converter/entities/omop/person?patient_uri={patient_uri_place_holder}
\end{lstlisting}

\begin{lstlisting}[float,frame=single,caption=Sample conversion rule from domain to application]
PREFIX schema: <http://schema.org/>
PREFIX func: <http://www.w3.org/2007/rif-builtin-function#>
PREFIX omop: <http://www.salusproject.eu/ontology/omop#>
PREFIX log: <http://www.w3.org/2000/10/swap/log#>
1  {	?person schema:birthDate ?birthdate.
2	(?birthdate) func:year-from-dateTime ?yearOfBirth.
3	(?birthdate) func:month-from-dateTime ?monthOfBirth.
4	(?birthdate) func:day-from-dateTime ?dayOfBirth.
5  } => {
6	?person omop:yearOfBirth ?yearOfBirth.
7	?person omop:monthOfBirth ?monthOfBirth.
8	?person omop:dayOfBirth ?dayOfBirth. }.
\end{lstlisting}

Listing 5 shows a conversion rule, which converts the birth date in the demographics domain entity to the OMOP person entity in the application domain. Line 2-4 use RIF builtins \cite{ polleres2009rif} to generate year, month and day from a birth date. The generated data is then converted to expressions using the OMOP ontology as displayed in Line 6-8. Listing 6 shows the sample data of the OMOP application entity, the data is generated by applying the conversion rule in Listing 5 to the domain data in Listing 4. As the predicates from the OMOP ontology used in Listing 6 are generated from the columns in the OMOP PERSON table with a 1:1 mapping, the ETL job to load the OMOP application entity to the OMOP database is simple and straight forward. Moreover, in our implementation, we applied SPARQL SELECT queries to each application entity to generate a CSV file, which further simplifies the ETL application. Table 2 shows a sample record in the PERSON table of the OMOP CDM database. The record in Table 2 is created by loading the sample application entity displayed in Listing 6.

\begin{lstlisting}[float, frame=single,caption=Sample application entity (Data Layer 4)]
<http://example.org/resource/Patient/1001> omop:yearOfBirth 1990 .
<http://example.org/resource/Patient/1001> omop:monthOfBirth 2 .
<http://example.org/resource/Patient/1001> omop:dayOfBirth 8 .
\end{lstlisting}

\begin{table}[ht]
\caption{Sample data of PERSON table in OMOP CDM database (Data Layer 5)}
\centering
\begin{tabular}{l l l l}
\hline
\textbf{person\_id} & \textbf{year\_of\_birth} & \textbf{month\_of\_birth} & \textbf{day\_of\_birth}\\
\hline
1001 & 1990 & 2 & 8 \\
\hline
\end{tabular}
\label{tab:omop-person}
\end{table}

\subsubsection{Proof of Conversion Process}
\label{sec:proof}

The previous sections demonstrate that using EYE as a semantic converter enables the conversion of RDF data over different data layers in a flexible and semantic way with explicit rules. It also shows that advanced functions are available with a list of builtins. Nonetheless there is still an important feature of the EYE reasoning engine to be introduced in this section: proof generation for the conversion process.

The EYE reasoning engine can generate a proof for its reasoning process. The proofs of the conversion processes discussed in Section 3.3 can be found online \cite{github-sample}. A proof records actions such as data extractions and inferences that lead to the conclusions of a reasoning process. The proof itself can be checked by third party N3 reasoning engines, e.g. CWM, to build trust on the reasoning process. In the data processing of the SDV the first step carried out by the RDF generator takes a one to one mapping, which does not change the semantics. The remaining conversion steps carried out by the semantic converter are able to provide proofs of the reasoning processes by EYE. Thus the entire semantic data processing is provable. Since the proofs can be checked independently, it gives valuable support for building trust needed by applications. We therefore claim that our SDV framework builds the foundation towards a full fledged implementation of the formalization layers presented in the Semantic Web Stack \cite{ bratt2007semantic}.

\section{Application and Performance}

Applications for semantically processing EHR data for clinical research, following the methods introduced in this paper, are implemented in several European projects. This section uses the SALUS project and the AP-HP project as examples to demonstrate the application and performance of the proposed semantic data virtualization solution.

\subsection{Application}
\label{sec:application}

\subsubsection{Semantic data processing in SALUS project}
\label{sec:salus}

The SALUS (Scalable, Standard based Interoperability Framework for Sustainable Proactive Post Market Safety Studies) project \cite{Salus} is an EU FP7 project which aims to create the necessary semantic interoperability infrastructure to enable secondary use of EHR data by various clinical tools for proactive post market safety studies. The semantic processing scheme presented in this paper is used in the SALUS project to build its semantic interoperability layer.

\begin{figure*}
\centering\includegraphics[width=0.64\linewidth]{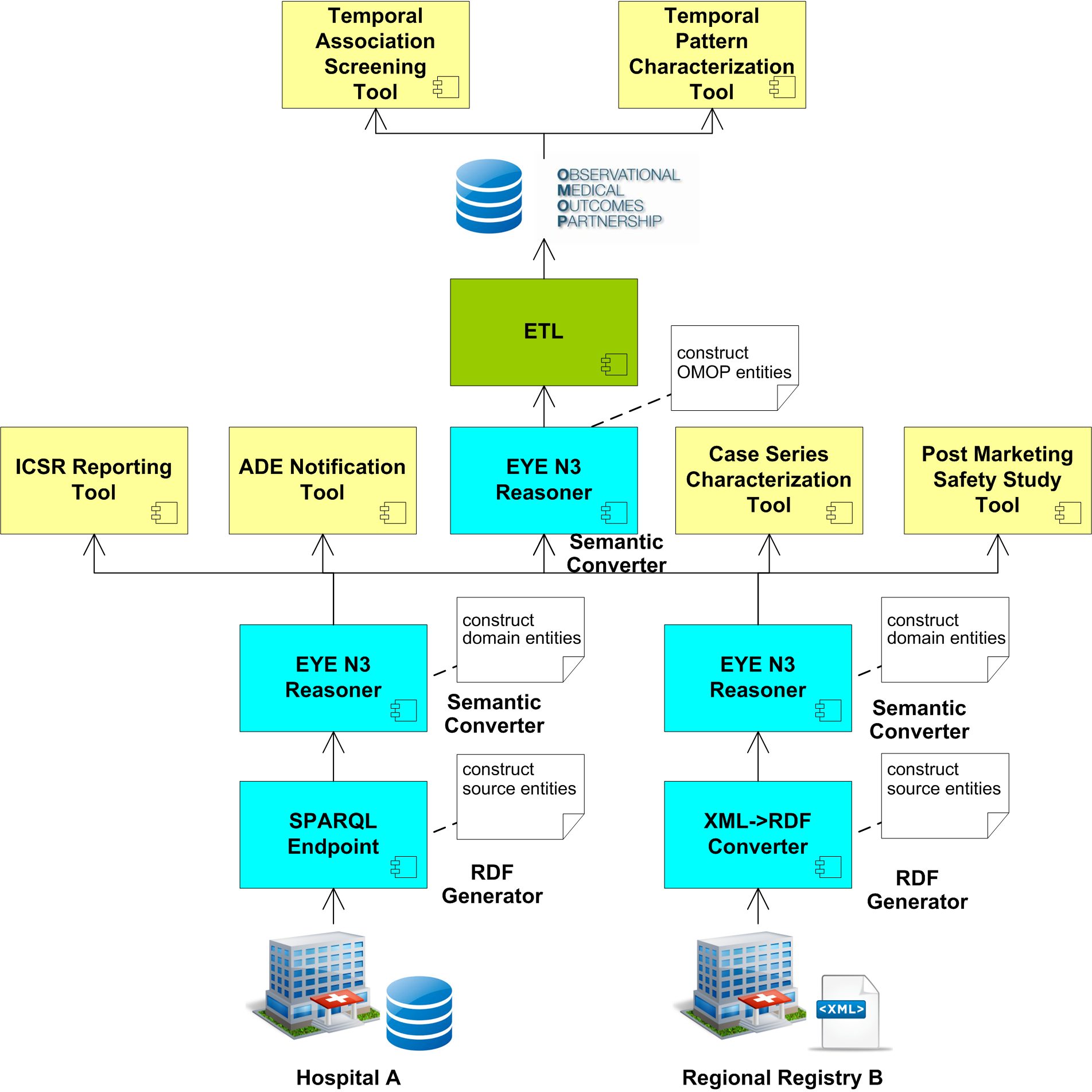}
\caption{Semantic data processing in SALUS}
\label{fig:salus-application}
\end{figure*}

Figure \ref{fig:salus-application} shows the data flow of the semantic data processing in the SALUS project. The source data of the SALUS project come from an EHR system of a large hospital and a large regional EHR registry. The EHR data in these two sources are stored in different formats: the records in a German Hospital A are stored in an Oracle database while the records in an Italian Regional Registry B are provided via an interface from an IHE QED profile in XML format. There are six clinical applications to support, which are divided into two sets: two temporal analysis related tools can only consume data from an OMOP database, the other four tools consume RDF data represented with CREAM \cite{CREAM}. 

A SPARQL endpoint and an XML to RDF converter are created for those two sources respectively, working as the RDF generator to construct source entities from the source data. The EYE reasoner is used as the semantic converter for both of the sources. Two different sets of conversion rules are created which converts data with each data source's semantics to harmonized domain semantics with CREAM. Terminology mappings, e.g. ICD 10 to SNOMED CT mappings, etc. are studied \cite{hussain2014justification} and manually created local lab code to LONIC mappings (with limited number) are also tested. However, due to the limited coverage of the existing mappings, as well as concerns of mapping qualities, code conversions are not carried out in the source to domain mapping process.

Those four tools that consume RDF data expressed with CREAM semantics can retrieve data from the EYE reasoner through the domain entity service, as we demonstrated in Section \ref{sec:example-do}. For the tools that retrieve data from the OMOP database, an extra conversion process is needed, which converts RDF data represented with CREAM semantics to representations with OMOP semantics. An OMOP ontology is created based on the database structure of the OMOP Common Data Model (CDM). In the SALUS project, only the demographics, diagnosis and medication entities are mapped to the corresponding OMOP application entities. The codes used in the diagnosis entity, namely ICD 9CM and ICD 10, and the codes used in the medication entity, namely ATC, are all contained in the OMOP vocabulary natively. Although OMOP provided a set of mappings to map ICD 9CM and ICD 10 codes to SNOMED CT, we did not carry out the coding conversion because the provided mappings only cover a limited set of the codes. An ETL process is executed to load the data to the OMOP database. The temporal analysis related tools are then able to consume data from the OMOP database.

Semantic interoperability as introduced in Figure \ref{fig:interoperability} is achieved in the implementation shown in Figure \ref{fig:salus-application}. A data source is able to provide data in different formats to support various clinical research applications and a clinical research tool is also able to use data from different data sources.

\subsubsection{Semantic data processing in AP-HP}
\label{sec:aphp}

The Assistance Publique – Hôpitaux de Paris (AP-HP) is the public hospital system of Paris and it is the largest hospital system in Europe. The semantic data processing scheme presented in this paper is applied in AP-HP to support advanced clinical applications. Experiments have been made to semantically process EHR data stored in AP-HP's ORBIS EHR system and load it to an i2b2 database for clinical research. Besides using i2b2, the semantically processed data is also loaded into SAS Visual Analytics to carry out operational and clinical analysis.

Both applications start with processing EHR data stored on data layer 1 to RDF representations on data layer 2, and later convert to harmonized domain entities on data layer 3 with CREAM. For the i2b2 experiment we created i2b2 ontologies based on the i2b2 database structure. The i2b2 platform uses a generic data model (star schema) as its database structure where a central fact table (OBSERVATION\_FACT) is joined with several dimension tables (e.g. PATIENT\_DIMENSION, CONCEPT\_DIMENSION, etc). The semantics of the facts stored in the OBSERVATION\_FACT table, e.g. whether it is a lab result or a diagnosis, is defined by the concept code (CONCEPT\_CD) column in the fact table. The meanings of the concept codes used in the facts are defined in the CONCEPT\_DIMENSION table, as well as the METADATA table in its ontology management cell \mbox{\cite{i2b2-ont}}.

When we created the application ontology for i2b2, we did not intend to cover the semantics that can be inferred from the generic data model of i2b2, e.g. whether a fact record is a lab result or diagnosis. Rather, we only create an ontology for each of the physical tables (fact, dimensions and metadata tables) by following a 1-1 mapping policy (see Section \mbox{\ref{sec:data-layer}}, Layer 2). Domain entities such as lab, diagnosis and procedure are all converted to an observation fact entity, differentiated by concept codes. The observation fact entity is loaded into the OBSERVATION\_FACT table by a lightweight ETL application. Terminologies used in the lab, diagnosis and procedures, e.g. CIM10 or CCAM codes, are loaded to the CONCEPT\_DIMENSION and METADATA tables. Application entities for the remaining dimension tables are also created and are loaded to their corresponding tables.

\begin{figure*}
\centering\includegraphics[width=0.95\linewidth]{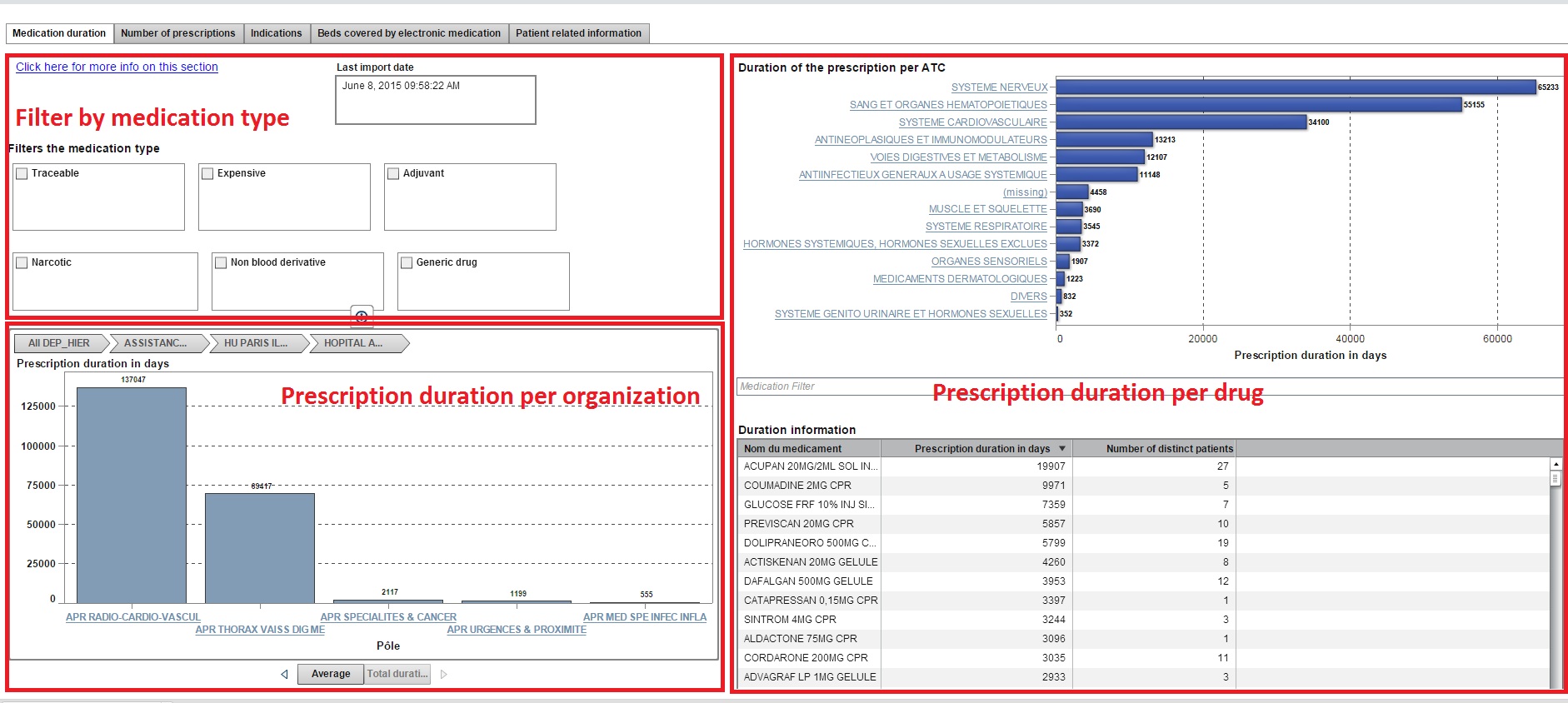}
\caption{Process data with SAS Visual Analytics}
\label{fig:sas}
\end{figure*}

The application for the SAS Visual Analytics is further differentiated from the OMOP and i2b2 application. Both OMOP and i2b2 applications require to create an application ontology and convert relevant domain entities to application entities before loading them into the target databases. The SAS application does not have a predefined database structure. It accepts input data as a CSV file and keeps the data as in-memory data sets. We therefore did not create application entities for SAS application, but directly load domain entities into the SAS VA application as in-memory data sets by a simple SAS data load job. Figure \mbox{\ref{fig:sas}} shows the analysis result in SAS Visual Analytics, based on the medication prescription domain entity.

\subsubsection{Application data flow}
\label{sec:data-flow}

\begin{figure*}
\centering\includegraphics[width=0.75\linewidth]{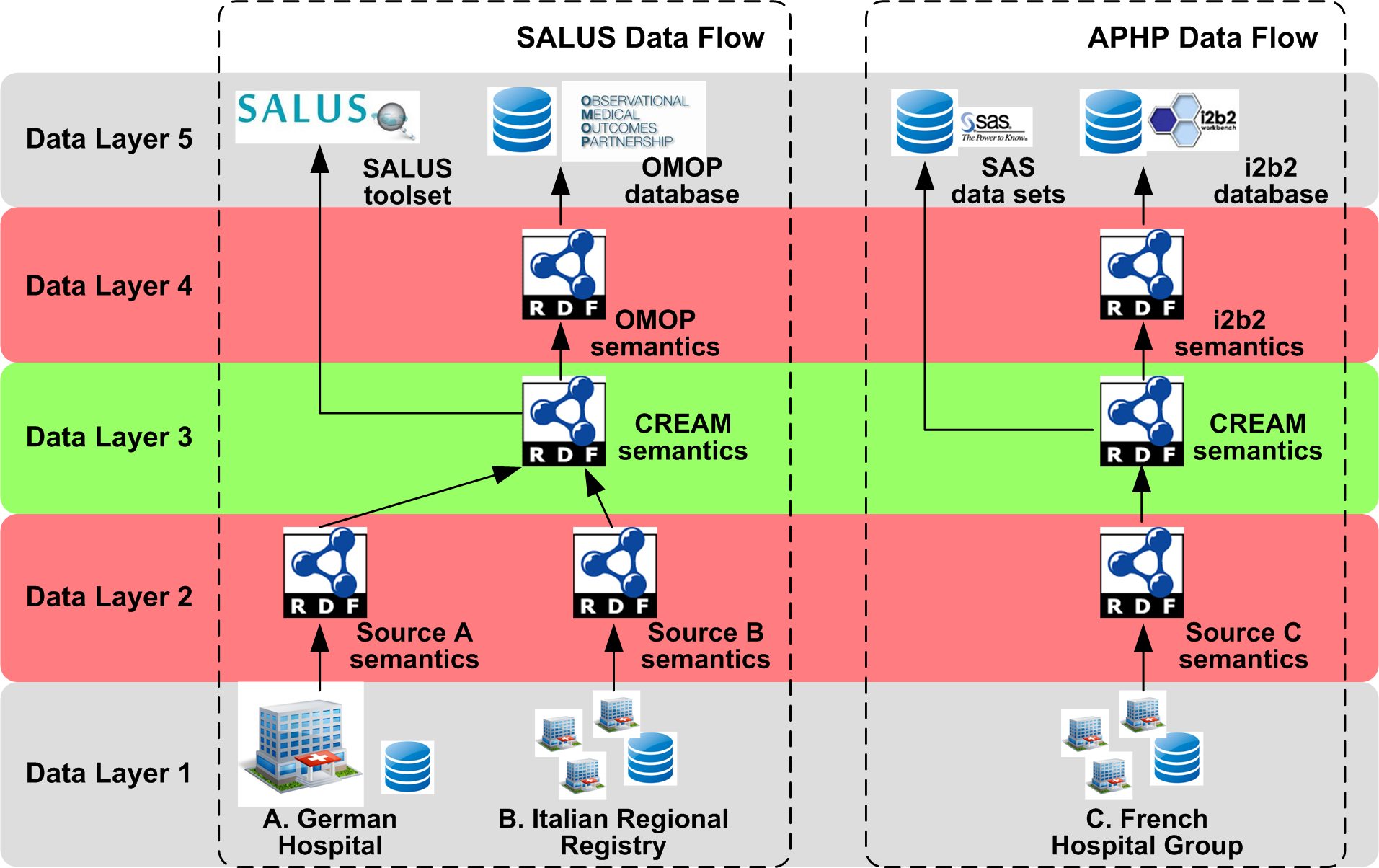}
\caption{Application data flow}
\label{fig:app-data-flow}
\end{figure*}

Figure \mbox{\ref{fig:app-data-flow}} shows the data flow in the aforementioned applications. It can be observed that EHR data from three independent clinical repositories with different sizes and located in three European countries are semantically processed and harmonized with CREAM semantics on data layer 3. In the SALUS project, the harmonized domain entities are further processed and loaded to an OMOP database to support temporal related analysis, meanwhile other SALUS tools consume domain entities directly. In the AP-HP project, the harmonized domain entities are further processed and loaded to the i2b2 database to support tailored clinical research applications, meanwhile the SAS Visual Analytics tool consumes domain entities directly. Since the EHR data in these two projects is both harmonized with CREAM on data layer 3, in principle semantic interoperability is achieved between the two projects. For example, the sample domain to OMOP conversion rule displayed in Listing 5 not only works on the demographics domain entity in the SALUS project, but also works on the demographics domain entity in the AP-HP project. We have not yet merged the data across the two projects due to data security constraints. However, it is expected that with minor modifications, the SALUS applications and the AP-HP applications can be adapted to work with domain data from both projects. Thus effectively integrating source A, B and C for different clinical applications as depicted in Figure \ref{fig:interoperability}.

\subsection{Performance}

The performance of the SDV is mainly defined by the efficiency of the RDF generator and semantic converter. Besides the software implementation, the performance of the RDF generator is also largely influenced by the speed of the data source. The performance of the semantic converter is largely influenced by the requested mapping, i.e. the complexity of the source and target pattern. 

The experiment described in this section is carried out in the pilot applications of the SALUS project. The pilot applications are implemented in a hospital that has 1,300 in-patient beds and 95 out-patient facilities. The data comes from a test database which contains anonymized electronic patient records from the production ORBIS EHR system. An ORBIS SPARQL endpoint is connected to the ORBIS database working as the RDF generator and the EYE reasoning engine works as the semantic converter. The semantic data virtualization solution is installed on a server which is equipped with a 2.0 GHz CPU with 8 cores and 32 GB memory. Six domain entities are constructed and implemented to support the applications listed in Section \ref{sec:application}. The implementation details of the SDV architecture introduced in Figure \ref{fig:salus-application} can be found in \cite{salus-d442}.

In this section, we show the performance on lab result and diagnosis entities. The lab result domain entity requests 2 source entities as input, and its conversion rule set contains 12 rules (each '=$>$' statement is counted as one rule). The diagnosis domain entity also requests 2 source entities as input, and its conversion rule set contains 9 rules. The test database contains 56 million lab result records and 13 million diagnosis records.

\begin{figure*}
\centering\includegraphics[width=0.95\linewidth]{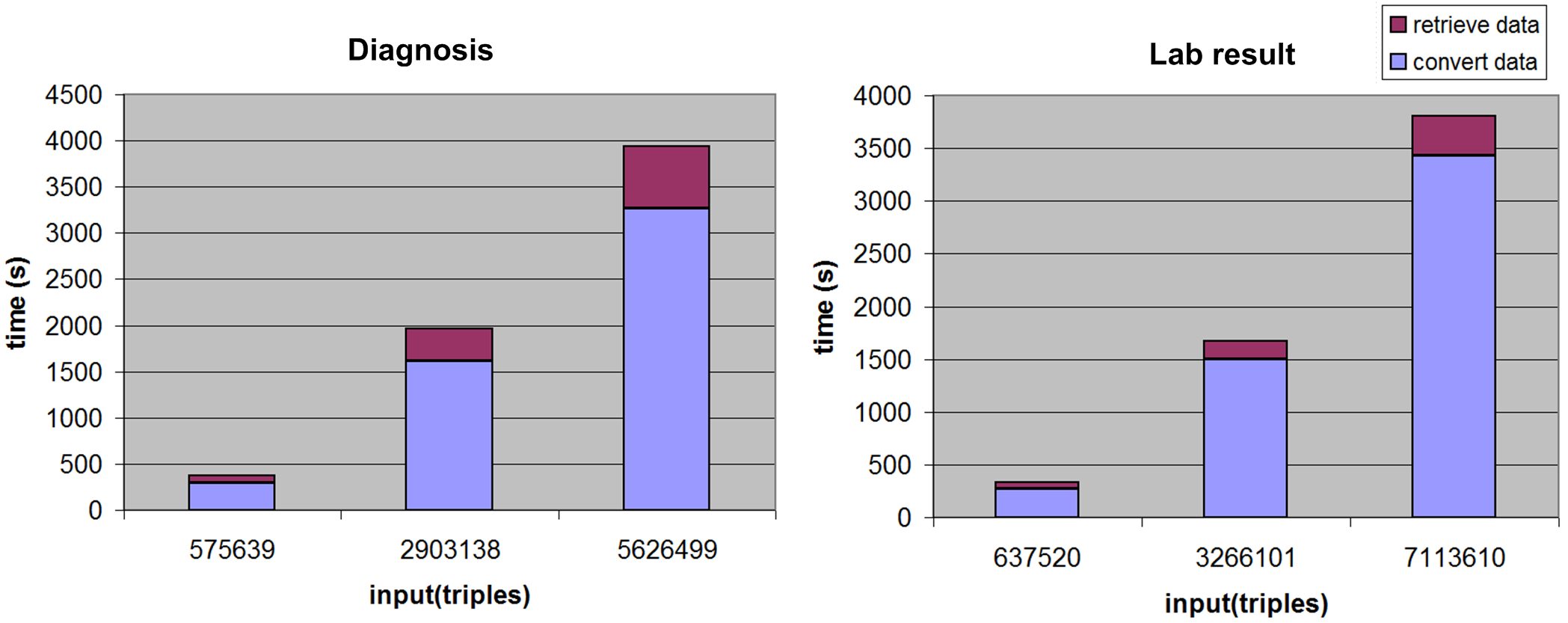}
\caption{Performance on domain entities}
\label{fig:Performance}
\end{figure*}

Figure \ref{fig:Performance} shows the performance of the diagnosis entity and the lab result entity. The retrieve data time refers to the time spent on the RDF generator to retrieve the requested source entities. The convert data time refers to the time spent on the semantic converter to translate source entities to a domain entity. The number of input triples indicates the size of the input source entities. Figure \ref{fig:Performance} shows that most of the time is spent on data conversion rather than data retrieval. 

The diagnosis test processes data recorded in a period of 1, 6, and 12 months respectively. There are 1.1 million diagnoses recorded in 12 months, which is equivalent to 5.6 million source triples. It takes 3,941 seconds to process those 1.1 million diagnoses, thus the process speed is 279 records/s. The lab result test processes data recorded in a period of 3, 14, and 31 days respectively. There are 2.6 million lab results recorded in 31 days, which is equivalent to 7.1 million source triples. It takes 3,809 seconds to process those 2.6 million lab results, thus the process speed is 682 records/s.

\begin{figure*}
\centering\includegraphics[width=0.7\linewidth]{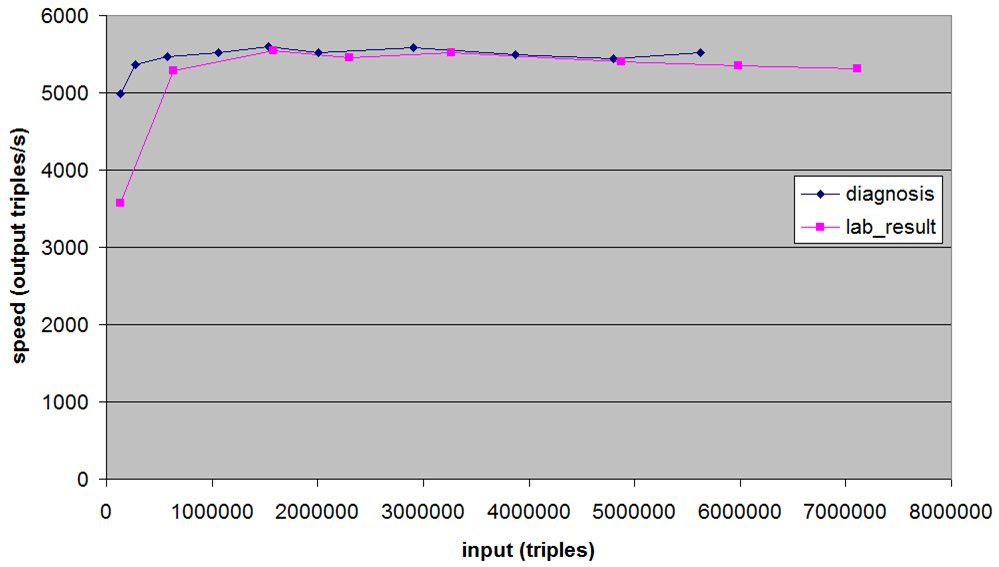}
\caption{Scalability test}
\label{fig:Scalability}
\end{figure*}

Figure \ref{fig:Scalability} tests the scalability regarding the speed of generating domain entities. The input axis indicates the size of the input source entities. The speed axis indicates how many triples, represented with domain semantics, are generated per second. The diagram shows that the speed is stable when the data size increases.

In practice, we are able to issue concurrent requests to retrieve the entities. For example, the diagnosis entity of year 2014 can be divided into 12 sub-entities, each one representing a month of the year. Those 12 sub-entities can be retrieved concurrently. Combining the results is straight forward because the sub-entities are independent and complementary. Given enough CPU cores, we are able to scale up. Nevertheless, the evaluation described in this section did not divide an entity into sub-entities.

\section{Conclusion and Future Work}
Semantic interoperability among different EHR systems and clinical research applications is becoming a rising demand for Healthcare IT systems. Standards are thus developed, aiming to regulate the data representation so as to improve interoperability among these different EHR systems. Nevertheless, it raises new challenges of interoperability between different standards.

We acknowledge the fact that there is no single standard that could serve as the interoperability hub for all the existing clinical data sources and clinical applications. This paper presents an approach to achieve semantic interoperability among different clinical data sources and different clinical applications by formalizing data with a semantic data virtualization (SDV) scheme. The SDV scheme allows producing multiple representations from a single data source to meet the requirements from different clinical applications.

The semantic data processing presented in this paper starts by formalizing the data sources with RDF representations, using the semantics of their respective database schemas. Semantic conversions are introduced in later steps, expressed as N3 rules. Such an approach preserves the provenance information of the data source, and makes the semantic conversion in an explicit and formal manner. Expressed as N3 rules, each semantic conversion step can provide proofs generated by an N3 reasoning engine. It is therefore able to construct proofs of the entire semantic conversion process, thus building trust on the generated results. We claim that our semantic interoperability framework builds the foundation for a working implementation of the layers presented in the Semantic Web Stack \cite{ bratt2007semantic}.

The software architecture of the proposed SDV solution is introduced. The SDV uses RESTful services to generate domain or application entities, which enables data transformation in a fully automated way. In addition, the virtualization policy supports run-time data transformation upon request, which avoids the burden of maintaining additional data stores to store processed data.

Examples of using the SDV to semantically process EHR data from disparate repositories and supporting different real-world clinical research applications are also presented. The interoperability between different data sources and clinical research applications are also discussed. The performance of the SDV on a clinical database is presented as well. To the best of our knowledge, this is the first report of a real-world application that semantically processes large volume clinical practice data in the entire data flow, where each semantic conversion step is explicitly expressed and provable. 

In future work, we will continue improving the scalability of the SDV. We intend to further investigate its ability to process large entities in memory, aiming to shift the capacity of in-memory entity sizes from millions of triples to billions of triples.

\section*{Acknowledgement}
This work was supported by funding from the SALUS project (http://www.salusproject.eu/). Grant agreement Number 287800. The authors would like to thank the anonymous reviewers for their valuable comments in improving the paper. The authors would also like to thank Els Lion for reviewing and improving the text of this paper.

\section*{Copyright}
The paper has been accepted for publication on Journal of Biomedical Informatics. Please retrieve a formal copy with DOI:10.1016/j.jbi.2015.10.009 (http://dx.doi.org/10.1016/j.jbi.2015.10.009)

© 2015. This manuscript version is made available under the CC-BY-NC-ND 4.0 license
http://creativecommons.org/licenses/by-nc-nd/4.0/



 \bibliographystyle{elsarticle-num} 
 \bibliography{ref}





\end{document}